\documentclass{ws-mpla}
\usepackage[super,compress]{cite}
\usepackage[breaklinks]{hyperref}  
\hypersetup{colorlinks,urlcolor=black,citecolor=black,linkcolor=black,filecolor=black}
\usepackage{graphicx}

\def\aap{Astron.\ Astrophys.\ }
\def\aaps{Astron.\ Astrophys.\ Ser.\ }

\def\prd{Phys.\ Rev.\ D\ }

\begin{document}

\markboth{Casanellas, J. and Lopes, I.}
{The Sun and Stars: Giving light to dark matter}

\catchline{}{}{}{}{}

\title{The Sun and Stars: Giving light to dark matter}

\author{\footnotesize JORDI CASANELLAS}
\address{Max Planck Institut f\"ur Gravitationsphysik (Albert-Einstein-Institut)\\
D-14476 Potsdam, Germany\\
jordi.casanellas@aei.mpg.de}

\author{IL\'IDIO LOPES}
\address{CENTRA, Instituto Superior T\'ecnico, Universidade de Lisboa\\
Av. Rovisco Pais 1, 1049-001 Lisboa, Portugal.\\
\medskip
Departamento de F\'isica, Universidade de Evora\\ 
Col\'egio Luis Ant\'onio Verney, 7002-554 Evora, Portugal\\
ilidio.lopes@tecnico.ulisboa.pt}

\maketitle

\pub{Received: 27 January 2014}{Accepted: 29 September 2014}

\begin{abstract}
During the last century, with the development of modern physics in such diverse fields as thermodynamics, 
statistical physics, and nuclear and particle physics, the basic principles of the evolution of stars have been 
successfully well understood. Nowadays, a precise diagnostic of the stellar interiors is possible with the new fields of
 helioseismology and  astroseismology.  Even the measurement of solar neutrino fluxes, once a problem 
in particle physics, is now a powerful probe of the core of the Sun. These tools have allowed the use of stars to test new physics, in particular the properties of the hypothetical particles that constitute the dark matter of the Universe. Here we present recent results obtained using this approach.
\end{abstract}

\ccode{PACS Nos.: 95.35.+d, 96.60.Ly, 97.10.Sj, 26.65.+t}

\section{Introduction} 

During the last 70 years, significant progress has been achieved in the study of the physics in the interior of the Sun and stars \cite{TurckChieze:1993dw,Catelan:2012pe,Guzik:2014boa}.
This advance has been possible due to the great development of several key fields of experimental and theoretical physics, 
 such as statistical physics, magneto-hydrodynamics, particle physics, and  nuclear physics, among others. 
Astronomers are now able to describe with high-precision the physics that takes place inside stars like the Sun,
as well as many other classes of stars, with masses different than that of the Sun,
and in quite different stages of stellar evolution. 
In general, the physics of stars is well understood since
the first moments of their formation, up to the most advanced stages of stellar evolution, including the formation of  
highly compact objects, like white dwarfs, neutron stars and black holes. The progress in the understanding of the physical principles operating inside stars was accompanied and challenged by important developments in the observational fields of astrophysics, 
such as astrometry, photometry and high-resolution spectroscopy, as well as the new fields
of helioseismology and asteroseismology, which create powerful tools to probe the interior of stars. 
The recent past has shown that the prosperity of modern stellar physics was possible due 
to the powerful partnership built between theoretical physics and astrophysics. 

\smallskip
In this new phase of stellar physics, the large amount of data made available by
several observational projects permits to use the Sun and stars as a tool to challenge our knowledge about fundamental physics,\cite{Vieira:2012pu} and in doing so it opens new branches of research, such as gravitation tests and probes of the existence of new particles. Among other applications, to validate the new gravitational theories proposed as an  alternative to General Relativity~\cite{2012ApJ...745...15C,2012PhRvD..85j4053A,2013PhRvD..87f1503S,2013PhRvD..88d4032H,2013arXiv1312.0705K}, 
and to probe the existence of dark matter (DM) inside stars to investigate the DM problem~\cite{2002MNRAS.331..361L,2012RAA....12.1107T}. 

\smallskip
Currently, the interest of studying the interaction of DM with stars is twofold: 
on one hand to identify which type of particles DM is made of, and, on the other hand,
to understand the physical mechanisms by which DM contributes to the formation of stars. 
In the former, stars are used as a complementary tool to test DM candidates, 
being in that way an alternative method to test the candidates proposed by modern theories of particle physics, or alternatively, to test candidates detected by experiments of direct or indirect DM searches. In the latter, the aim is to explore how DM contributes for the structure formation in the Universe, comprising galaxies and the first generation of stars, not only by locally changing the gravitational field where stars are formed, but also to explore how the interaction of DM with baryons changes the evolution of stars.
   
\section{\sc Present status of the dark matter problem}

Our Universe is constituted by 5\% of {\it baryonic matter}, a type of matter in which we have become great experts during the last two centuries, by developing several branches of Physics; 27\% is constituted by {\it dark matter}, which plays a major role in the formation of structure in the Universe, but its fundamental characteristics are 
yet poorly understood; and another 68\% of the total energy density of the Universe is usually refereed to as {\it dark energy},
 which physical origin is even more uncertain~\cite{Ade:2013zuv}. Although the basic properties of DM are not known, namely, which type
 of particles is DM made of, there is strong evidence of its existence, 
both from astrophysical and cosmological observations, as well as from numerical simulations~\cite{2005PhR...405..279B,2013PhST..158a4014B}. 
Among other direct evidence of the existence of a gravitational field caused by the presence of DM, 
we make reference to the velocity of galaxies in clusters, the rotation curves of galaxies, 
the cosmic microwave background anisotropies, the velocity dispersions of dwarf spheroidal galaxies and 
the inference of the DM by gravitational lensing~\cite{2008ARA&A..46..385F}.  
All these observational and theoretical results suggest that most of the formation of structure in our Universe 
can only be explained by the presence of a 
gravitational field caused by the presence of a new type of particles
that must be non-baryonic and cold~\cite{Munshi:2011jg}, such as the particles belonging to the group of the WIMPs (for weakly interacting massive particles). 
Therefore, it is no surprise that with such an amount of observational evidence for the existence of DM, a large effort is being devoted to theoretical work and experimentation in several branches of astrophysics, cosmology and particle physics, with the intent of discovering such fundamental particles. If DM particles exists, then in the near future we
should expect to detect such particles in the Large Hadron Collider at CERN or in 
other direct detection experiments. Another possibility is the confirmation of
the existence of DM by the indirect detection of DM by-products, 
like the production of high-energy neutrinos or gamma rays caused by the annihilation of DM pairs. 

\smallskip
In the last 30 years, several classes of particles have been proposed as DM candidates, 
among others,  axions, WIMPs, asymmetric DM particles, and other more 
exotic types of matter.\cite{Taoso:2007qk,2010ARA&A..48..495F}
At present, two groups of particles merit special attention, 
because they sum up most of the  critical properties necessary  to be the ideal DM particle. 
First, the well known Weakly Interacting Massive Particles (WIMPs), which  
interact gravitationally with other particles and have weak interaction  with  baryons. 
WIMPs are among  the most popular DM candidates. Such class of particles  
occur in several extensions of the Standard Model of particle physics, 
like super-symmetric (SUSY) models.\cite{1996PhR...267..195J}  
In such models, the lightest SUSY particle, the neutralino, 
a stable particle with a self-annihilation cross section of the order of the  weak-scale interaction, 
is the most suitable candidate for DM.
The second type of candidates are known as asymmetric DM 
particles,\cite{2005PhLB..605..228H,2006PhRvD..73k5003G,Khlopov:2007ic,2008PhRvD..78f5040K,2009PhRvD..80c7702F,Khlopov:2010pq,Davoudiasl:2011fj,Petraki:2013wwa,Zurek:2013wia,Kumar:2013vba,2014ApJ...780L..15L} 
which like WIMPs have interactions with baryons at the weak-scale,  
even if they do not self-annihilate inside compact objects.\cite{2009PhRvD..79k5016K} Unlike WIMPs, 
these particles carry a conserved charge analogous to the baryon number asymmetry. As a consequence, DM
 becomes asymmetric, \textit{i.e.}, 
there is an unbalanced amount of particles and antiparticles, 
introducing an asymmetric parameter in the DM sector identical 
to baryon-anti-baryon asymmetry parameter, the so-called baryonic asymmetry. 
Furthermore, these particles are expected to have 
a mass of the order of a few GeV~\cite{2011arXiv1102.5644K,2010PhRvD..82e6001C}.

\smallskip
In recent years, several underground experiments  have been built  
to search for direct signatures of the interactions between DM particles and a baryons~\cite{2005PhR...405..279B}. 
Most of such DM searches have not detected any DM signal. This is the case
of experiments like XENON10/100,\cite{2011PhRvL.107e1301A,Aprile:2013doa} PICASSO,\cite{Archambault:2012pm} SIMPLE,\cite{Felizardo:2011uw} and LUX.\cite{Akerib:2013tjd}  However, in disagreement with such results are 
the positive detections of DAMA/LIBRA,\cite{Bernabei:2010mq} CoGeNT,\cite{2011PhRvL.107n1301A} CRESST-II~\cite{Angloher:2011uu} and CDMS II~\cite{Agnese:2013rvf} experiments,
which all found evidence of events that can be credited to DM particles with similar properties. The former two experiments report evidence 
of an annual modulation in the differential event rate, which is explained 
as a consequence of the motion of the Earth around the Sun, which in turn moves through a cloud of DM particles~\cite{1986PhRvD..33.3495D}. Nevertheless, these results remain utterly controversial and further 
experimental work must be done in order to converge on a plausible answer about these detections.   

\smallskip
A possible solution to accommodate all experimental results comes from a new theoretical 
interpretation of the interaction of DM particles with baryons in the detectors. 
The simpler interpretation, 
used so far to analyse the data, is to assign 
the  annual modulation to a collision of a DM particle with nucleons inside the detector. In such scenario, the DM particle is estimated 
to have a  mass of the order of a few GeV (likely between  $5$ and $12$ GeV),  
and a DM-nucleon scattering cross-section of the order of $10^{-40}\;\textmd{cm}^2$ for spin-independent (SI) interactions or $10^{-36}\;\textmd{cm}^2$ for spin-dependent (SD) interactions on protons. 
DAMA and CoGeNT experiments show very similar positive detections, yet the DAMA experiment is 
favourable  to a larger scattering cross-section than the CoGeNT experiment.
Nevertheless, this problem is resolved by several theoretical solutions that
have been proposed to overcome the inconsistency between the different experimental 
results~\cite{2010JCAP...08..018C,Hooper:2010uy,2011NuPhB.853..607F,2011PhRvD..84h3001H,2011PhRvD..84b7301D,Savage:2010tg}. Under such interpretations the data obtained by the different experiments can be reconciled.

\smallskip
Among the various possible theoretical explanations, one of the more appealing suggestions is the possibility that the DM particle couples unequally to the protons and neutrons of the collision  
nuclei because of isospin violation. Usually, the  DM particle is considered to couple equally 
with protons and neutrons. Accordingly, the SI scattering cross-section of heavy elements 
scales with $A^2$, where $A$ is the atomic number of the nucleus.
If the DM particles couple differently to protons and neutrons, that leads to a quite 
distinct interpretation of  direct DM searches, 
leading to the reconciliation of almost all the experiments~\cite{2010JCAP...08..018C,2011PhRvD..84b7301D}. 
This nuclear mechanism is known as isospin 
coupling violation~\cite{2004PhRvD..69f3503K,2005PhRvL..95j1301G,2009NJPh...11j5026C,2012PhRvD..85h1301K}.
In such cases, the proton scattering cross section for the SI interaction increases 
by $10^2$ to $10^3$ relatively to the usual interpretation of the experimental results, leading to an effective SI scattering cross-section 
with values between $10^{-40} {\rm cm^2}$ and $10^{-36} {\rm cm^2}$.\cite{2011PhRvD..84b7301D,2011PhRvL.107n1301A}. Other models propose momentum and velocity-dependent interactions to explain the data.\cite{Chang:2009yt,Fitzpatrick:2012ix} 
We point out here that these types of DM particles can modify the internal properties of stars like the Sun by conducting energy very efficiently,\cite{Vincent:2013lua} leading to a quite different flux of solar neutrinos~\cite{2010Sci...330..462L}
and helioseismology data~\cite{2010PhRvD..82h3509T,2010ApJ...722L..95L}.  

\section{\sc The physics of dark matter particles inside stars}

Dark matter impacts the evolution of a star by means of two mechanisms:  
through the change of the transport of energy \cite{1985ApJ...294..663S,Gould:1989ez} and by the creation of an additional source of energy.\cite{1989ApJ...338...24S} The former mechanism can even be important for stars in our galactic neighbourhood, like the Sun~\cite{DeRujula:1985wm,Dearborn:1990mm} and other main-sequence stars.\cite{Casanellas:2012jp} The latter mechanism is more pronounced in environments with a very high DM density, several million times the DM density of the solar neighbourhood. This type of scenario occurs in  stellar populations located in the centre of galaxies, including the Milk Way,\cite{2009MNRAS.394...82S} or during the formation of the first generation of stars in the primordial Universe.\cite{Spolyar:2007qv} On the other hand, the gravitational influence of the DM mass accumulated in the interior of stars similar to the Sun is several tens of orders of magnitude smaller than the star's total mass, so it can be neglected. 

\smallskip
Most DM particles cross the stellar interior without undergoing any type of interaction, yet a few of them scatter off nuclei losing part of their kinetic energy. In some cases, the loss of energy will result
in the particle being captured by the star, \textit{i.e.}, the particle can no more escape the star's gravitational field.
As expected,  a star with a large mass has a large gravitational field, and consequently captures a larger amount of DM during its evolution.
The capture rate of DM particles by a star is inversely proportional 
to the mass and the dispersion velocity of the DM particle,
but proportional to the local DM density of the halo and
the scattering cross-section off baryons~\cite{1987ApJ...321..571G}. 
Two leading parameters define the scattering of the DM
particles with the nuclei of the stellar isotopes: the SD scattering cross-section that
is only relevant for hydrogen; and the SI scattering cross-section
that defines the interaction of the DM particles with the heavy nuclei.  
The values of the scattering cross sections used in the calculations with stellar codes are usually the maximum allowed by the null results of direct detection experiments, or those in agreement with the values suggested by the interpretation of positive experimental results. However, it is worth noticing that, if the value of the SI scattering cross-section is larger 
than a hundreth of the SD 
scattering cross-section, then the capture of DM particles is dominated by collisions 
with heavy nuclei, rather than by collisions with hydrogen~\cite{2011PhRvD..83f3521L}.
The scenario that is usually considered corresponds to a DM halo with a density in the solar neighbourhood of $0.3 \; {\rm GeV\; cm^{-3}}$, constituted by particles with a mass of a few GeV and with a Maxwellian velocity distribution with a dispersion of $270 \; {\rm km\; s^{-1}}$.

\smallskip
The interactions of DM particles with stars have been implemented in a stellar code~\cite{art-Morel1997} that explicitly follows the capture rate of the DM particles by 
the different chemical elements present inside stars, some of them changing  
in  isotopic abundance during the star's evolution.
Capture rates of DM were first calculated by Ref.~\refcite{1985ApJ...296..679P} in the case of the Sun, by Ref.~\refcite{1987ApJ...321..571G} for generic massive bodies, and by Ref.~\refcite{1989ApJ...346..284B} for main sequence stars.
Presently, the capture rate is computed numerically from the integral expression of Ref.~\refcite{1987ApJ...321..571G} 
implemented as indicated in Ref.~\refcite{2004JCAP...07..008G}. 
A detailed discussion about the dependence of the capture on the properties of DM particles, as well as the impact that the uncertainties in the DM and stellar parameters have on the capture rate by the Sun and other stars is presented in Ref.~\refcite{2011PhRvD..83f3521L}.

\smallskip
The accumulation of DM inside a star leads to the formation of a very small DM core, with a radius of the order of the few percent of the stellar radius. The impact on the stellar properties comes from the additional energy transport mechanism provided by DM conduction, or by the DM self-annihilation. In both cases, the evolution of the star is only affected when the changes produced by the DM particles compete with the classical transport or production
of energy mechanisms. Mostly, only low-mass stars can be influenced, and the impact of DM becomes insignificant for more massive stars.

\smallskip
The efficiency of the energy transport provided by DM particles
depends on the average distance travelled by  particles between 
consecutive collisions, \textit{i.e.}, the mean free path of the DM particle. 
If the mean free path is short compared with the DM scale height, 
then the stellar plasma and the DM are in local thermal equilibrium. Alternatively, if the mean free path is large, 
the successive collisions are widely separated,
so that the energy transfer proceeds within the Knudsen regime 
\cite{2002MNRAS.337.1179L}. 

\smallskip
The energy generation rate due to pair annihilation  of DM  particles is more
significant in DM halos of high density.
In such cases, the DM in the stellar core provides an extra
source of energy. It follows that every pair of captured DM particles annihilates, 
being converted into additional energy for the star. Without loss of generality,
all the products of DM annihilation, except neutrinos, are assumed to interact with the stellar plasma, so these new particles have a very short mean free paths in the star's core and rapidly reach the thermal equilibrium. 
The process can be very efficient: for each annihilation pair most of the energy is converted in thermal energy and only a small fraction is lost  
in the form of high-energy neutrinos that escape the star's gravitational field. 
Recent simulations of the self-annihilation of two neutralinos (which are Majorana particles in most SUSY models) have shown that the energy loss in a star as the Sun could be as low as 10\% of the total energy produced by the annihilation of the DM pair~\cite{2009MNRAS.394...82S}.
\smallskip

\section{Impact of DM particles in stars and constraints on the DM properties}
\begin{figure}
\begin{tabular}[]{cc}
\hspace{-0.8cm} \includegraphics[scale=0.38]{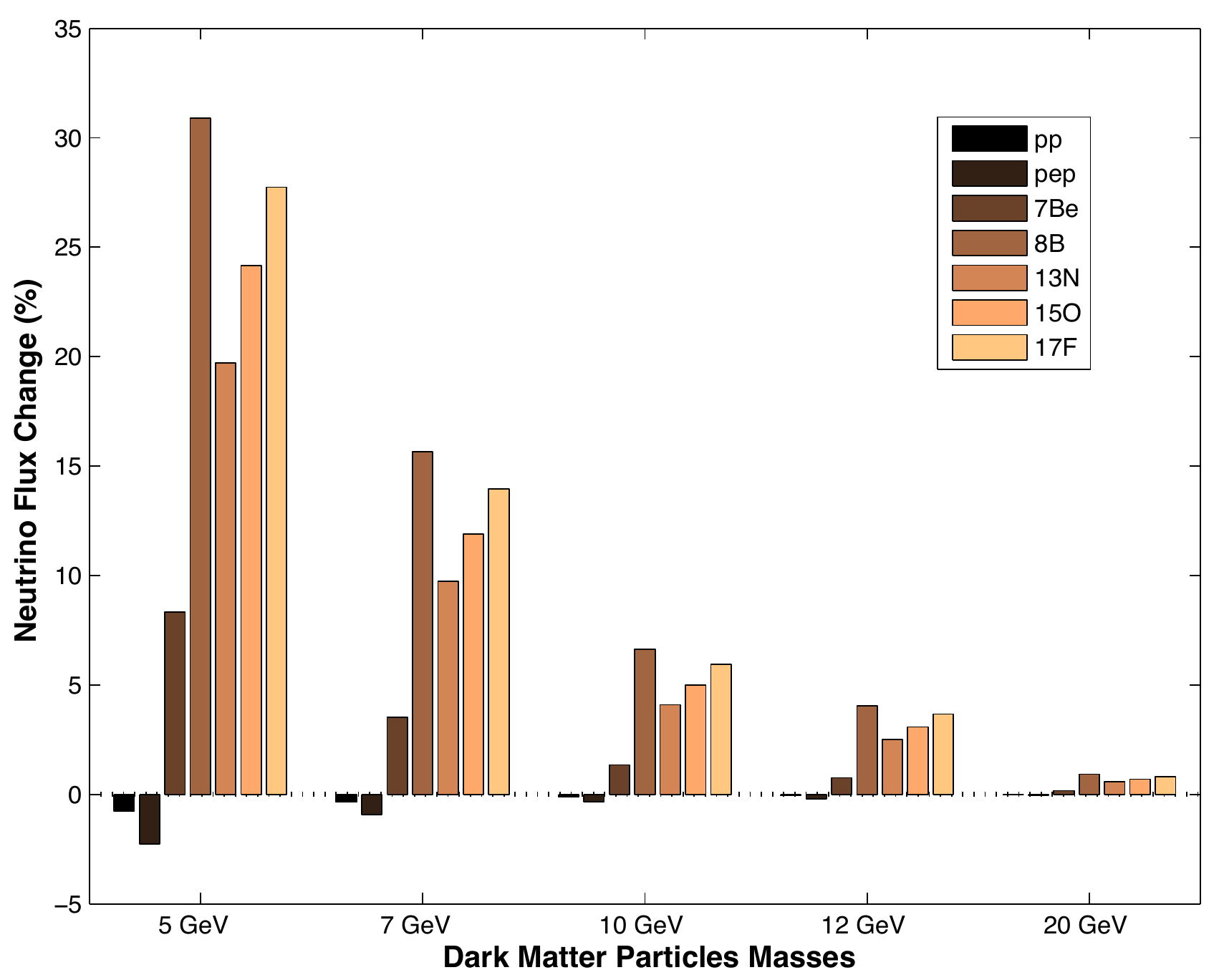}  &
\includegraphics[scale=0.36]{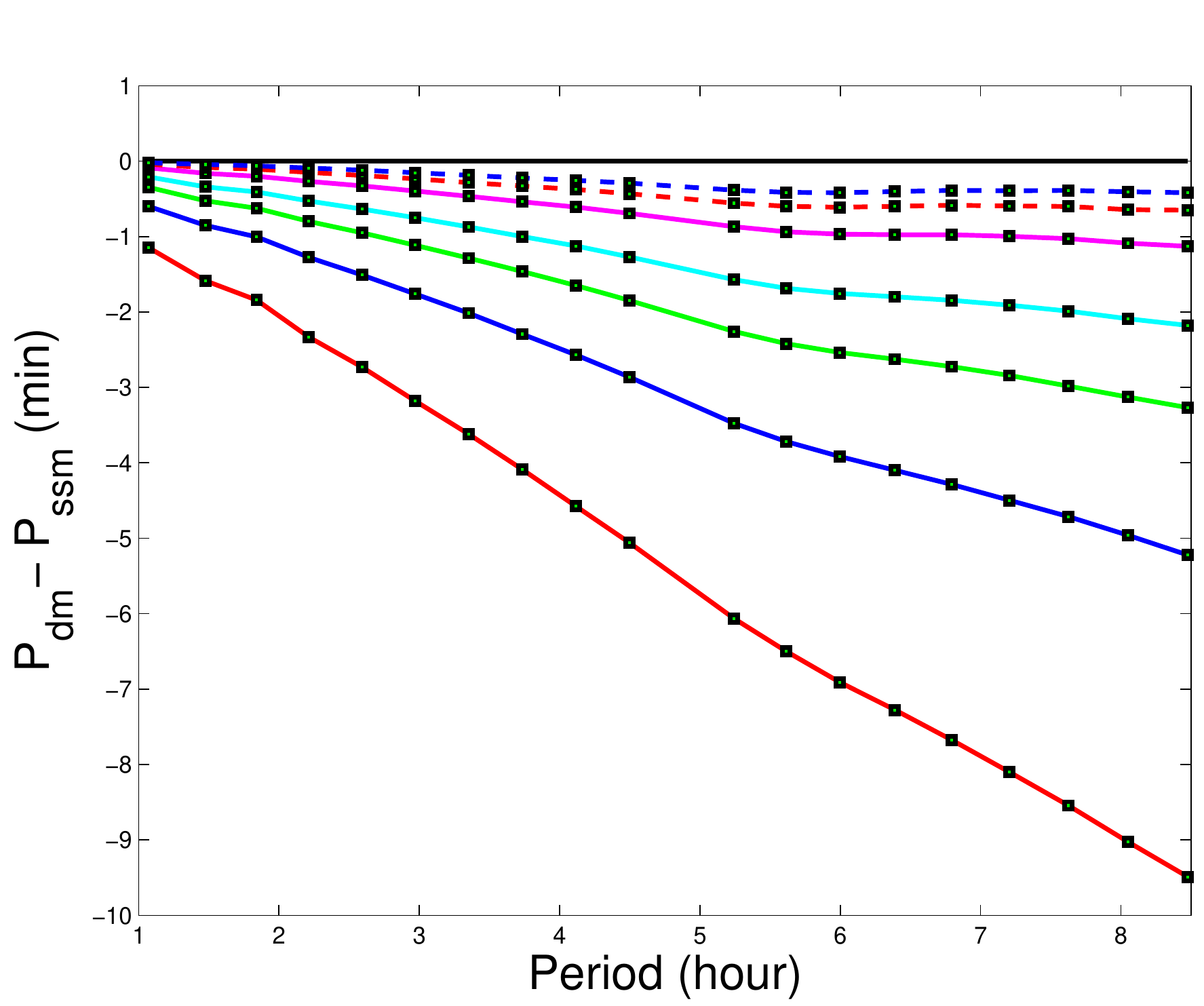} 
 \end{tabular}
 \caption{Comparison of the predicted neutrino fluxes (left figure) and the periods $P_{l,n}$ of the dipole gravity modes $l=1$ (right figure), between the solar standard model and models of the Sun that include the influence of DM particles with different characteristics. See Refs.~\cite{2010Sci...330..462L,2010ApJ...722L..95L} for the details.}
\label{fig-neut_helio}
\end{figure}

Probing DM by determining its impact on the evolution of the Sun and stars is possible under the 
strong hypothesis that the physics inside the stars is known within the required accuracy. The computation of sophisticated modified solar models have shown that a region of the DM parameter space can be ruled out in the case of asymmetric DM candidates or, more generally, when the DM particles do not annihilate after being captured, so they accumulate in larger numbers. The primary effect on the structure of the star is to provide an additional mechanism for the transfer of energy, which in some cases reduces significantly the temperature gradient of the plasma, thereby causing the stellar core to become almost isothermal.\cite{2002PhRvL..88o1303L,2010ApJ...722L..95L} The existence of asymmetric DM particles with small mass and large scattering cross sections off baryons would imply strong modifications on the solar neutrino fluxes (see Figure~\ref{fig-neut_helio}.left) and changes in the internal structure~\cite{2014ApJ...780L..15L} 
 that would be detected by helioseismology (see Figure~\ref{fig-neut_helio}.right). Solar neutrino fluxes have been also shown to be useful to probe the parameter space of isospin-violating DM candidates~\cite{2012ApJ...752..129L,2012ApJ...757..130L}.
 \smallskip

\begin{figure}
\centering
\includegraphics[scale=0.8]{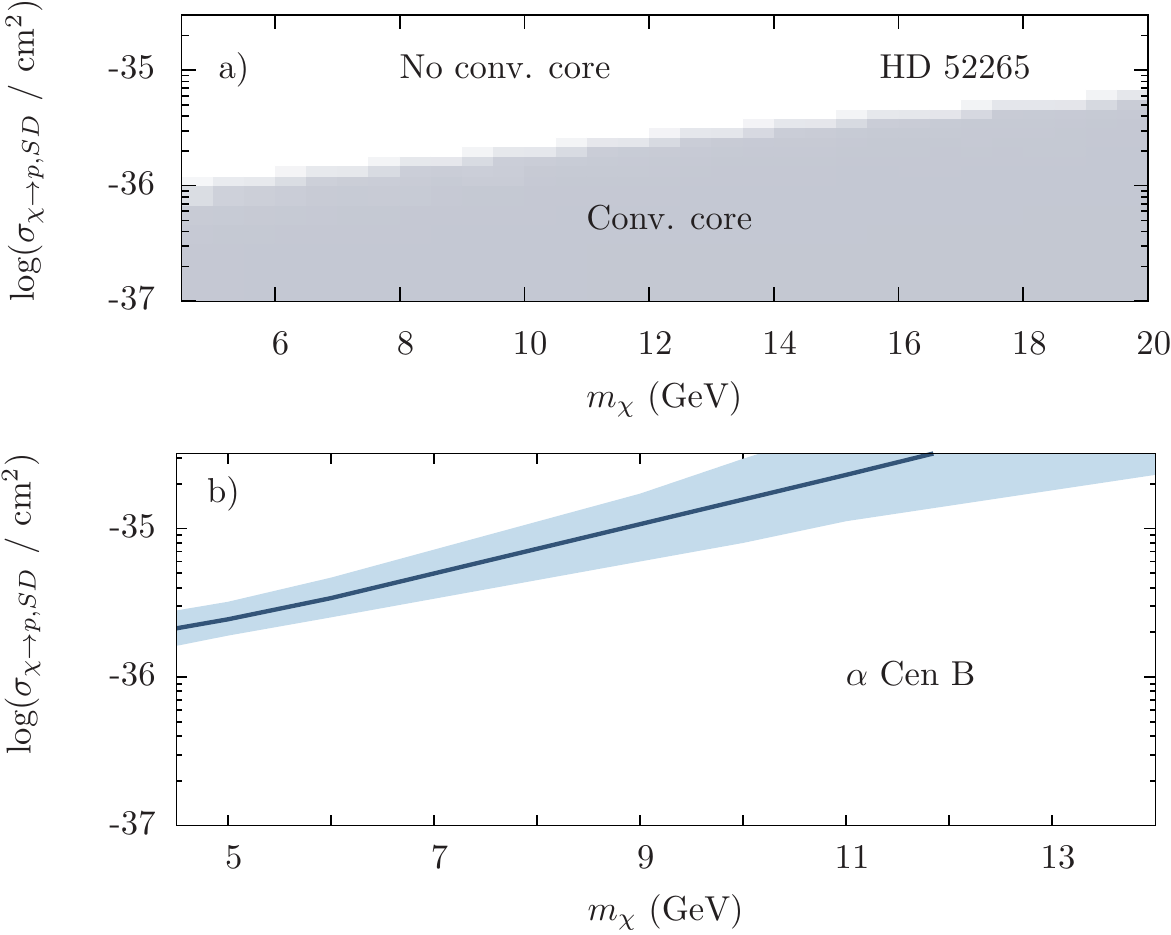}
\caption{a) The convective core predicted by the classical theory of stellar evolution in the star of 1.2 M$_{\odot}$ HD 52265 is not expected when the DM particles have large scattering cross section with protons, $\sigma_{\chi\rightarrow p,SD}$, and a small mass, $m_{\chi}$. b) The changes in the central properties of the star $\alpha$~Cen B due to the presence of asymmetric DM could be detected in the acoustic oscillations of the star. Asymmetric DM particles with properties above the blue line are ruled out because they lead to a small separation between the frequencies of the oscillation modes of low degree that are more than 2 $\sigma$ away from the observations. The blue region around the line shows the uncertanty on the limits when the observational errors are taken into account. In both cases an environmental DM density of 0.4 GeV cm$^{-3}$ was assumed. See Ref.~\cite{Casanellas:2012jp} for more details.}
\label{fig-multi-ADM}
\end{figure}

Similar modifications due to the existence of asymmetric DM are also expected in nearby stars. The motivation to explore the effects of DM in very low-mass stars comes from the fact that the lower the mass of a star, the more strongly it will be influenced by the accumulation of DM.\cite{Zentner:2011wx,Iocco:2012wk} In addition, other main sequence stars with masses within a particular range can have their internal structure importantly modified: it has been shown that the convective core expected in 1.1-1.3 M$_{\odot}$ stars would disappear depending on the properties of the DM particles (see Figure~\ref{fig-multi-ADM}.a).

\smallskip
Taking advantadge of the fact that the nearest star to the Sun, $\alpha$~Cen, is a binary system formed by two stars which characteristics are known with a high precision, the authors in Ref.\refcite{Casanellas:2012jp} were able to constrain the parameter space of asymmetric DM studying the oscillations of $\alpha$~Cen B. Scattering cross sections above 10$^{-36}\;\textmd{cm}^2$ for a DM particle mass of 5~GeV were ruled out using this approach. The constraints, shown in Figure~\ref{fig-multi-ADM}.b), were imposed comparing the results of modified stellar models with the observed small separations between the frequencies of the acoustic modes of oscillation with low degree, $\langle\delta \nu_{02}\rangle$. The asteroseismic parameter $\langle\delta \nu_{02}\rangle$ is sensitive to the properties of the core of the star, where the modifications induced by the accumulation of DM occur.
\smallskip

Asymmetric DM can also potentially produce huge effects in compact stars,\cite{Bertone:2007ae,2012JCAP...10..031L,Leung:2013pra} such as the creation of a small black hole inside a neutron star that may eventually destroy the star. \cite{Kouvaris:2011fi,Bramante:2013hn,2013PhRvD..87l3507B,Bramante:2013nma,Kouvaris:2013kra}
\begin{figure}[!t]
\centering
\begin{tabular}[]{cc}
\hspace{-0.6cm}
 \includegraphics[scale=0.6]{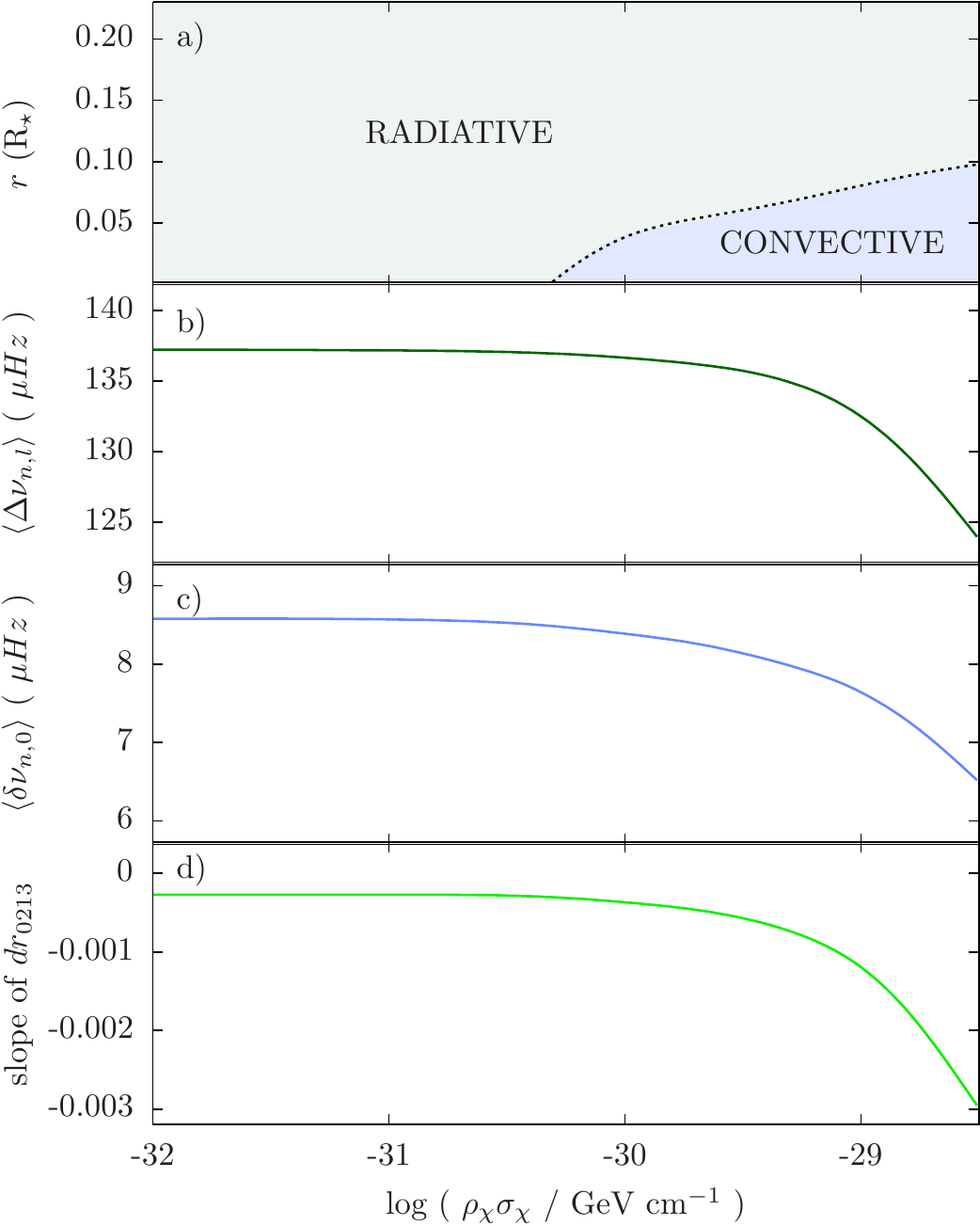} &
 \includegraphics[scale=0.8]{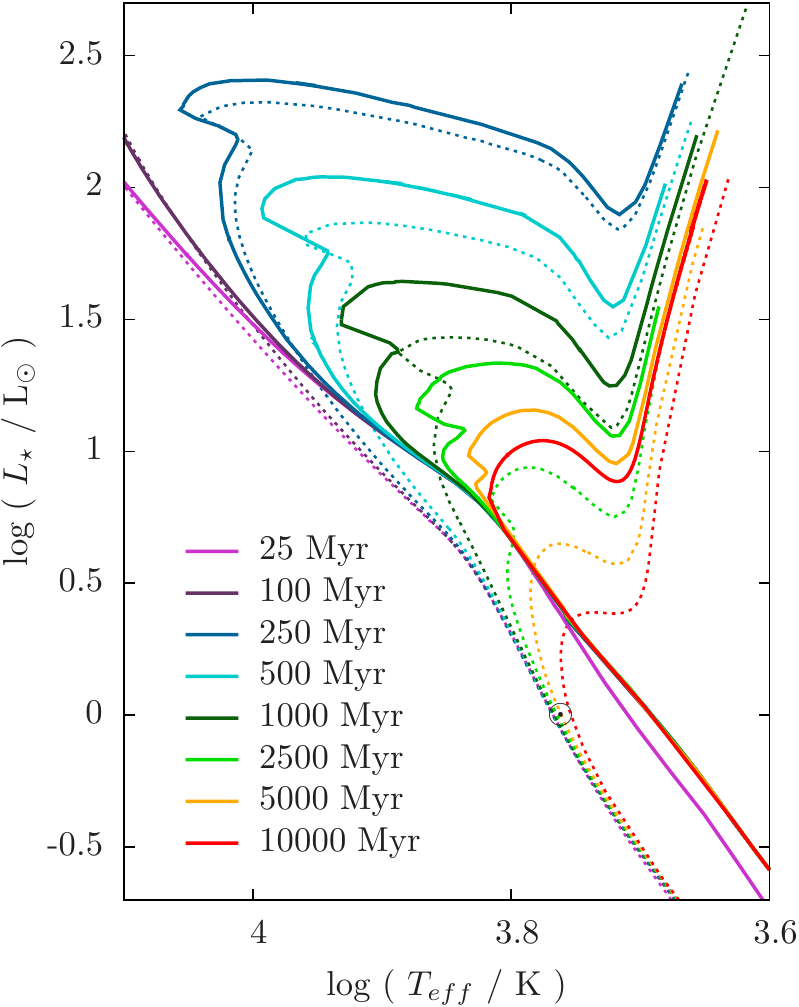}
 \end{tabular}
\caption{Left: (a) Size of the convective core, and the calculated seismologic parameters: (b) mean  large separation, (c) mean small separation and (d) slope of $dr_{0213}$, for 1$\;$M$_{\odot}$ stars that evolved in DM haloes with different densities $\rho_{\chi}$ and SD WIMP-nucleon cross-sections $\sigma_{\chi,SD}$, when the stars reached a luminosity $L=1\;$L$_{\odot}$.
See Ref.~\cite{2011MNRAS.410..535C} for the details. 
Right: Isochrones for a cluster of stars with masses between 0.7 M$_{\odot}$-3.5 M$_{\odot}$ that evolved in a halo of DM with a density
 $\rho_{\chi}=10^{10}\;$GeV cm$^{-3}$ (continuous lines) and for the same cluster in the classical scenario without DM (dashed lines).
See Ref.~\cite{2011ApJ...733L..51C} for the details.}
\label{fig-seism_isoc}
\end{figure}

\smallskip
In other scenarios, the DM annihilation provides stars with an extra source of energy that can dramatically change their evolution path.\cite{Moskalenko:2007ak,2009MNRAS.394...82S} However, environmental DM densities millions of times greater than the local DM density are needed for these important modifications to occur. These huge DM densities are only expected in very particular locations in the local Universe, such as the galactic center or the dwarf spheroidal galaxies, and in the early stages of the Universe. The study of the impacts of DM in stars in environments with high DM densities is subjected to important difficulties from the observational side and larger uncertainties from the theoretical side. The most important ones are those on the stellar mass and velocity, as well as on the density of DM around the star, which strongly depends on the model. On the other hand, in these cases the potential effects of DM on low-mass stars can be dramatic: the creation of an unexpected convective core 
that may be 
detected with asteroseismology 
(see Figure~\ref{fig-seism_isoc} left),\cite{2011MNRAS.410..535C} the evolution through a different path in the Hertzsprung-Russell diagram at a lower evolutionary speed,\cite{2009ApJ...705..135C} and the changes in the global properties of a whole cluster of stars (see Figure~\ref{fig-seism_isoc} right).\cite{2011ApJ...733L..51C} Similarly, the first generation  of stars may have also been strongly influenced by the presence of DM~\cite{Ripamonti:2010ab,Gondolo:2010kq,Zackrisson:2010jd,Scott:2011ni,2012MNRAS.tmp.2794I,Smith:2012ng,Stacy:2013xwa}.
\smallskip

\section{Conclusion}

In the last fifteen years, the use of the Sun and stars as cosmological tools to probe and test the DM particle 
candidates has become a regular procedure in DM research. This is testified by the arrival of several 
groups working in such a new research field~\cite{2010PhRvD..82j3503C,2010PhRvD..82h3509T,2011arXiv1110.1169H}, as well as by a significant
 increase on the number of publications in this subject. The increase of our knowledge of the physics of the solar interior made possible by the solar neutrinos as well as by helioseismolgy will allow accurate tests of the existence of different DM particle candidates, and also the extension of such studies to other fields of fundamental physics, such as alternative theories of gravitation.
 \smallskip
 
Presently, with the large amount of data made available by observational asteroseismology, mostly by the spacial missions Kepler\cite{Gilliland:2010tx} and CoRoT,\cite{2009A&A...506..411A} the pulsation spectrum has been identified and measured in more than ten thousand stars. Some of these stars are identical to the Sun, but most of them have quite different masses and they are in distinct phases of stellar evolution. With this large and diverse set of stars it should be possible to further constrain the properties of the DM particles.
\smallskip

The approach highlighted here provides a complementary contribution to the multidisciplinary effort of DM research. Only a collaborative work across the several fields involved will allow the scientific community to produce fruitful results, and in doing so to overcome one of 
the most exciting and difficult problems of modern astrophysics, particle physics and cosmology, which is the discovery of the constituent particle(s) of DM. 
 
\section*{Acknowledgements}
J.C. acknowledges the support from the Alexander von Humboldt Foundation. I.L. acknowledges the support from the Funda\c c\~ao para a Ci\^encia e Tecnologia.

\bibliographystyle{ws-mpla}

\end{document}